\documentclass[eqsecnum,aps,prb,twocolumn]{revtex4}

\usepackage{graphicx}%
\usepackage{dcolumn}
\usepackage{amsmath}

\makeatletter
\def\btt#1{\texttt{\@backslashchar#1}}%
\DeclareRobustCommand\bblash{\btt{\@backslashchar}}%
\makeatother

\begin{document}

\title{
Exotic Heavy Fermion State in the Filled Skutterudite PrFe$_4$P$_{12}$\\
Uncovered by the de Haas-van Alphen Effect\\
}

\author{
H.~Sugawara,~$^{1}$ T.~D.~Matsuda,~$^{1}$ K.~Abe,~$^{1}$ Y.~Aoki,~$^{1}$ H.~Sato,~$^{1}$ S.~Nojiri,~$^{2}$ Y.~Inada,~$^{2}$ R.~Settai,~$^{2}$ and Y.~\={O}nuki~$^{2,3}$\\
}

\affiliation{
$^1$Graduate~School~of~Science,~Tokyo~Metropolitan~University,Minami-Ohsawa,~Hachioji,~Tokyo~192-0397,~Japan\\
$^2$Graduate School of Science, Osaka University, Toyonaka, Osaka, 560-0043, Japan\\
$^3$Advanced Science Research Center, Japan Atomic Energy Research Institute, Tokai, Ibaraki 319-1195, Japan
}

\date{\today}

\begin{abstract}
~~~We report the de Haas-van Alphen (dHvA) experiment on the filled skutterudite PrFe$_4$P$_{12}$ exhibiting apparent Kondo-like behaviors in the transport and thermal properties. We have found enormously enhanced cyclotron effective mass $m^{\rm \ast}_{\rm c}=81~{\it m}_{\rm 0}$ in the high field phase (HFP), which indicates that PrFe$_4$P$_{12}$ is the first Pr-compound in which really heavy mass has been unambiguously confirmed. Also in the low field non-magnetic ordered phase (LOP), we observed the dHvA branch with $m^{\rm \ast}_{\rm c}=10~m_{0}$ that is quite heavy taking into account its small Fermi surface volume (0.15\% of the Brillouin zone size). The insensitivity of mass in LOP against the magnetic field suggests that the quadrupolar interaction plays a main role both in the mass renormalization and the LOP formation.\\

\end{abstract}


\maketitle

\section{INTRODUCTION}
Heavy fermion (HF) materials in strongly correlated $f$-electron systems have attracted much attention over the last two decades because of their interesting physical properties.~\cite{Stewart} Until now, numerous rare earth HF compounds have been found, however, most of them are trivalent Ce($f^1$)- or Yb($f^{13}$)-based systems with Kramers doublet ground states. There have been intense experimental researches to investigate new exotic phenomena in $f^{2}$ systems mainly on U-compounds. One of interesting characteristics predicted for $f^{2}$ systems is the quadrupolar Kondo phenomena theoretically proposed by Cox to explain the non-fermi-liquid behaviors observed in several U-compounds with non-magnetic crystal electric field (CEF) ground state.~\cite{Cox} However, most of U-compounds are not suitable to be described by the CEF scheme because of the highly itinerant nature of 5$f$-electrons. On the contrary, for most Pr-compounds, strong hybridization effect is hardly expected due to the localized character of 4$f$-electrons. In fact, Kondo-like behaviors have been observed in limited Pr-compounds, such as PrSn$_{3}$ and PrInAg$_{2}$.~\cite{Settai,Yatskar} For the latter, really large Sommerferd coefficient of $\gamma\sim 6.5$ J/K$^{2}\cdot$mol, comparable with those reported for the typical Ce- or U-heavy fermion compounds, has been reported. However, no direct evidence of the highly enhanced effective mass has been obtained, since the sample with high enough quality to detect the dHvA signals has not been prepared.

\begin{figure}[!tbp]
\begin{center}\leavevmode
\includegraphics[width=0.85\linewidth]{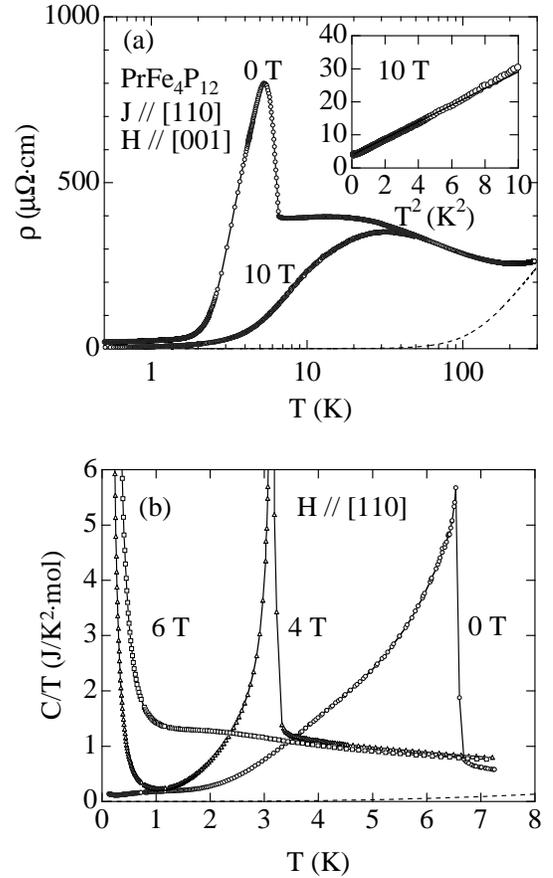}
\caption{Temperature dependence of (a) the electrical resistivity $\rho$  and  (b) specific heat divided by temperature $C/T$ in  PrFe$_4$P$_{12}$. The dashed lines in both figures are the phonon contribution estimated from LaFe$_4$P$_{12}$ data.}
\label{r&C/T}
\end{center}
\end{figure}

Ternary intermetallic compounds RT$_{4}$X$_{12}$ (R = rare earth; T = Fe, Ru, and Os; X = P, As, and Sb) with the filled skutterudite structure (Im\={3}) show a rich variety of electrical and magnetic properties depending on the components R, T, and X.~\cite{Meisner1981,Torikachvili1985,Sekine} Among them, PrFe$_4$P$_{12}$ is the most interesting compound because of the apparent Kondo-like behaviors found in the transport properties and specific heat $C$ which are not expected for the well localized 4$f$-electron system.~\cite{Sato_PRB,Matsuda,localized_chara} Figure~\ref{r&C/T} shows the temperature $T$ dependence of (a) electrical resistivity $\rho$ and (b) specific heat divided by temperature $C/T$.
The apparent $-{\rm log}T$ dependence of $\rho$ above $\sim$30~K and also the large magnitude of thermoelectric power at low $T$ ($-70~\mu{\rm V/K}$ above 6.5~K) suggest dominant contribution from the Kondo effect.~\cite{Sato_PRB} Existence of highly correlated electrons is manifest in the behaviors of $\rho$ and $C/T$ in the high field state above the metamagnetic transition field $H_{\rm M}\sim$5.5~T. Under 10~T, $\rho$ follows $\rho_{\rm 0}+AT^{\rm 2}$ [see the inset in Fig.~\ref{r&C/T} (a)] below $\sim$2~K with a large value of $A=2.5~\mu\Omega\cdot$cm/K$^2$. 
In the specific heat measurement under 6~T, $C/T$ increases with decreasing $T$ and saturates below $\sim$2~K to an anomalously large value of $\sim 1.4$~J/K$^{2}\cdot$mol.
An upturn at lower $T$ is the nuclear Schottky contribution $C_{\rm n}$ mostly due to Pr nuclei. By subtracting $C_{\rm n}/T$, we have estimated the electronic contribution $C_{\rm el}/T(\equiv\gamma)$ at 0~K as 0.10, 0.13, and $1.2$~J/K$^{2}\cdot$mol under 0, 4, and 6~T, respectively. 
Note that the large $\gamma$-value above $H_{\rm M}$ along with the $A$-constant follows the Kadowaki-Woods relation indicating a typical HF behavior.~\cite{Kadowaki} It is of interest to clarify the origin of the unusual HF behaviors in PrFe$_4$P$_{12}$.

Another anomalous nature is the phase transition at 6.5~K below which the Hall effect measurement revealed a carrier reduction indicating the Fermi surface (FS) reconstruction.~\cite{Sato_PRB} The transition was first ascribed to the superzone gap formation associated with an antiferromagnetic (AF) transition.~\cite{Torikachvili1985} However, in the recent neutron scattering experiment,~\cite{Keller} no Bragg peak expected for an AF ordering has been detected below 6.5~K, suggesting the ordered state to be non-magnetic, consistent with the non-magnetic CEF ground state (either $\Gamma_{1}$ or $\Gamma_{3}$) inferred from the anisotropy in the magnetization [$M(H\|\langle 100 \rangle) > M(H\|\langle 110 \rangle) > M(H\|\langle 111 \rangle)$].~\cite{Y_Aoki} That is further supported by the small upper bound of the Pr-magnetic moment ($\le 0.03\mu_{\rm B}$) estimated from the small Pr-nuclear contribution to the specific heat under 0~T.~\cite{Y_Aoki}
The softening of the elastic constants $C_{11}$ and ($C_{11}-C_{12})/2$ below $\sim$30~K reported in the recent ultrasonic measurement also suggests the CEF ground state to be non-magnetic $\Gamma_{3}$ and the phase transition to be an antiferro-quadrupolar ordering.~\cite{Nakanishi}

All these facts point to the possibility of quadrupolar Kondo effect in PrFe$_{4}$P$_{12}$ competing with the quadrupolar ordering. The dHvA experiment can be the most powerful tool to clarify the unusual HF state in this exotic Pr-compound; to obtain direct evidence of highly enhanced effective mass and to study most precisely the electronic structure such as FS. In this paper, we report the first successful dHvA effect both in the low field ordered phase (LOP) and the high field phase (HFP). The preliminary report for the dHvA experiments have been published.~\cite{Sugawara_JMMM}

\section{EXPERIMENTAL}
Single crystals of PrFe$_4$P$_{12}$ and LaFe$_4$P$_{12}$ were grown by the tin-flux method basically same as described in ref.6. The raw materials were 4N (99.99\% pure)-Pr, -La, -Fe, 6N-P, and 5N-Sn. The crystal structure was verified by the X-ray powder diffraction measurement. The lattice constant $a=7.815$~\AA~is close to the reported value.~\cite{Meisner1981}
The dHvA experiments were performed in a 17 T superconducting magnet with a top loading dilution refrigerator system cooled down to 28 mK. The dHvA signals were detected by means of the conventional field modulation method with a low frequency ($\sim10$~Hz).

\section{RESULTS AND DISCUSSION}
\begin{figure}[!tbp]
\begin{center}\leavevmode
\includegraphics[width=0.85\linewidth]{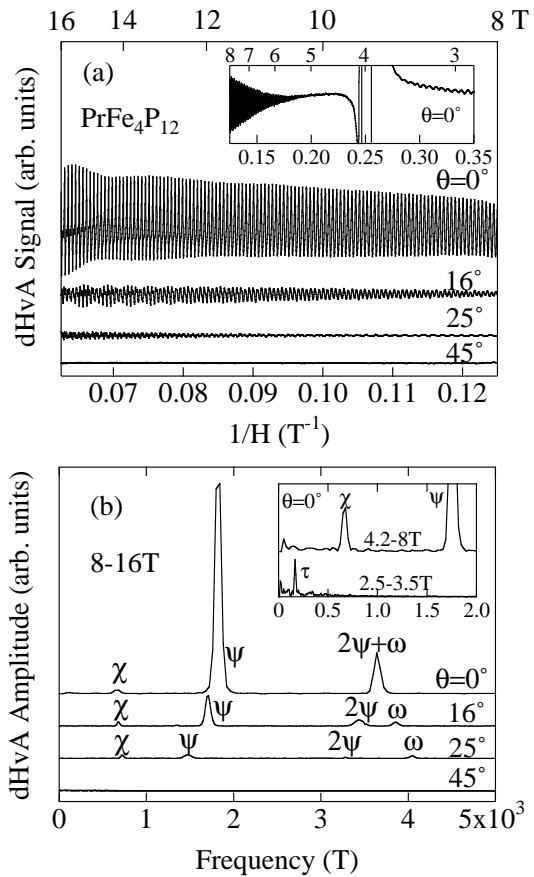}
\caption{(a) Typical recorder traces of the dHvA oscillations and (b) the corresponding Fourier spectra in PrFe$_4$P$_{12}$. $\omega$, $\psi$, $\chi$, and $\tau$ are fundamental dHvA frequency branches and 2$\psi$ is the 2nd harmonic of $\psi$. A remarkable feature is a reduction of the dHvA signal amplitude with increasing $\theta$. The insets show the
evolution of oscillations of the $\psi$- and $\chi$-branches and the vanishing of the $\tau$-branch in the HFP.}
\label{PrFe4P12dHvA_Osc&FFT}
\end{center}
\end{figure}

Figure~\ref{PrFe4P12dHvA_Osc&FFT} shows (a) typical recorder traces of the dHvA oscillations for selected field directions and (b) corresponding Fourier spectra, where $\theta$ is a tilting field angle from [001] to [101] in (010).
The three dHvA branches $\omega$, $\psi$, and $\chi$ were observed only above $H_{\rm M}$, while the branch $\tau$ was observed only below $H_{\rm M}$ (see the insets), which implies a large change in FS across $H_{\rm M}$.

\begin{figure}[!tbp]
\begin{center}\leavevmode
\includegraphics[width=0.85\linewidth]{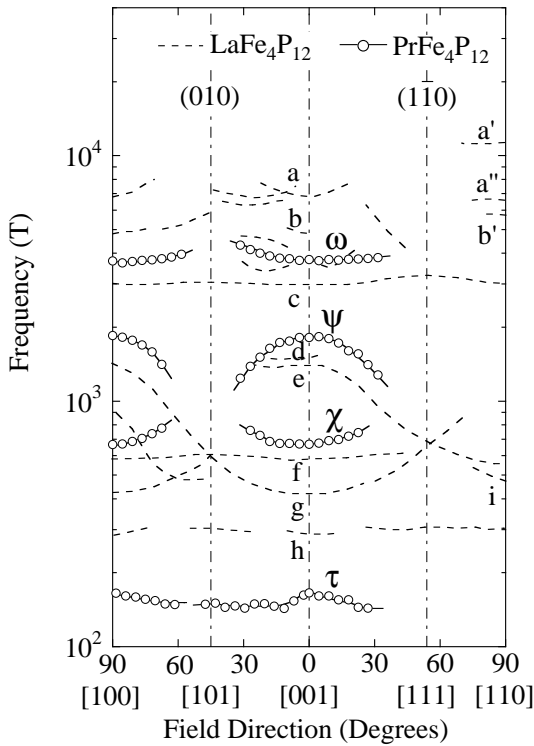}
\caption{Comparison of the angular dependence of the dHvA frequencies between PrFe$_4$P$_{12}$ (circles) and LaFe$_4$P$_{12}$ (dashed lines).}
\label{PrFe4P12dHvA_ang}
\end{center}
\end{figure}

Figure~\ref{PrFe4P12dHvA_ang} shows the angular dependence of dHvA frequencies in PrFe$_4$P$_{12}$ along with that in LaFe$_4$P$_{12}$.
The comparison of dHvA frequencies and $m^{\rm \ast}_{\rm c}$ between these compounds is given in Table~\ref{table1}.~\cite{Table1}
\begin{table}
\caption{Comparison of the typical dHvA frequencies $F$ and the cyclotron effective masses $m_{\rm c}^*$ between PrFe$_4$P$_{12}$ and LaFe$_4$P$_{12}$.}
\begin{ruledtabular}
\begin{tabular}{@{\hspace{\tabcolsep}\extracolsep{\fill}}ccccc}
Branch & $F(\times10^3$ T) & $m_{\rm c}^*$($m_0$) & $F(\times10^3$ T) & $m_{\rm c}^*$($m_0$)   \\ \hline 
\multicolumn{1}{c}{PrFe$_4$P$_{12}$}  & \multicolumn{2}{c}{$\theta=0^\circ(H\|\langle 100 \rangle$)} & \multicolumn{2}{c}{$\theta=25^\circ$} \\
   $\omega$ & 3.86\rlap{*}  & 62\rlap{*}   & 4.08  & 65 \\
        $\psi$ & 1.85  & 9.7  & 1.45  & 34 \\
        $\chi$ & 0.67  & 32  & 0.74  & 42 \\
        $\tau$ & 0.17  & 10  & 0.15  & 10 \\ 
\hline
\multicolumn{1}{c}{LaFe$_4$P$_{12}$}  & \multicolumn{2}{c}{$\theta=0^\circ(H\|\langle 100 \rangle$)} & \multicolumn{2}{c}{$\theta=45^\circ(H\|\langle 110 \rangle$)} \\ 
    a, a\rlap{''} & 6.91  &  6.8  & 6.61 & 22 \\
    b      &4.85   & 8.2  &   &  \\
    c   & 3.00  & 2.4  & 3.04   &  2.3  \\
    e   & 1.41  & 6.7  &    &   \\
    g   & 0.42  & 3.3  & 0.60   &  8.8  \\
\end{tabular}
\begin{flushleft}
 *$\theta=16^\circ$; the frequency for $\omega$-branch is too close to separate from the 2nd harmonic of $\psi$-branch, then we could not estimate $m_{\rm c}^*$ for $\omega$-branch at $\theta=0^\circ$.
\end{flushleft}
\label{table1}
\end{ruledtabular}
\end{table}
The branches $\omega$, $\psi$, and $\chi$ in PrFe$_4$P$_{12}$ are observed in the limited angular range centered at $\langle 100 \rangle$, and could not be detected around $\langle 110 \rangle$ and $\langle 111 \rangle$. In contrast, $\tau$-branch is observed in the whole angular range in $\{ 100 \}$. Assuming a spherical FS for this branch (the volume of FS is only 0.15\% of the 1st Brillouin zone size), the carrier number is estimated to be $1.3\times10^{19}/{\rm cm}^{3}$ which is close to $+1.9\times10^{19}/{\rm cm}^{3}$ estimated from the Hall coefficient.~\cite{Sato_PRB} The rough agreement between the estimated carrier numbers could be understood, since lighter carriers (holes in the present case) give dominant contribution to the Hall coefficient in the low fields. 
The drastic reduction of the $\gamma$-value in the LOP is explained by the carrier reduction associated with the FS reconstruction. However, the Sommerfeld coefficient of $\gamma$=5~mJ/K$^{2}\cdot$mol estimated from the $\tau$-branch assuming a spherical FS with 10~$m_{\rm 0}$ is still smaller than the experimental $\gamma\sim$100~mJ/K$^{2}\cdot$mol. According to the nesting model,~\cite{Sugawara_JPSJ,Harima} PrFe$_4$P$_{12}$ loses a large part of FS and becomes a semimetal in LOP. This scenario is consistent with the antiferro-quadrupolar transition associated with the FS instability with a nesting vectors ${\bf q}=$~(1,0,0).~\cite{Curnoe,Iwasa} There should be heavier electron-like FS(s) ($m^{\rm \ast}_{\rm c}\sim200~m_{\rm 0}$) that compensates the observed hole-like FS and dominates
the observed $\gamma$-value in LOP.

The most important findings from the dHvA experiments can be summarized as follows.

{\it 1)~Enormously enhanced effective mass}

\begin{figure}[!tbp]
\begin{center}\leavevmode
\includegraphics[width=0.85\linewidth]{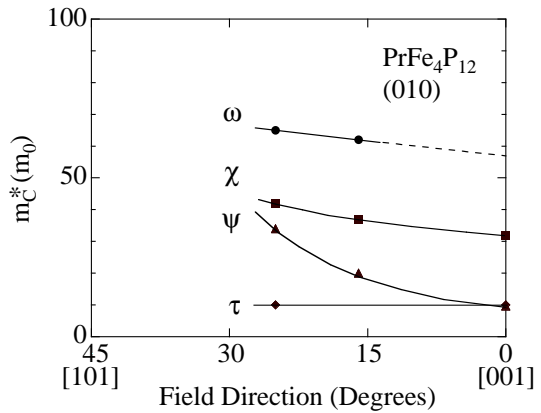}
\caption{Angular dependence of the cyclotron effective masses $m_{\rm c}^*$ in PrFe$_4$P$_{12}$.}
\label{AngDepMass}
\end{center}
\end{figure}

\begin{figure}[!tbp]
\begin{center}\leavevmode
\includegraphics[width=0.85\linewidth]{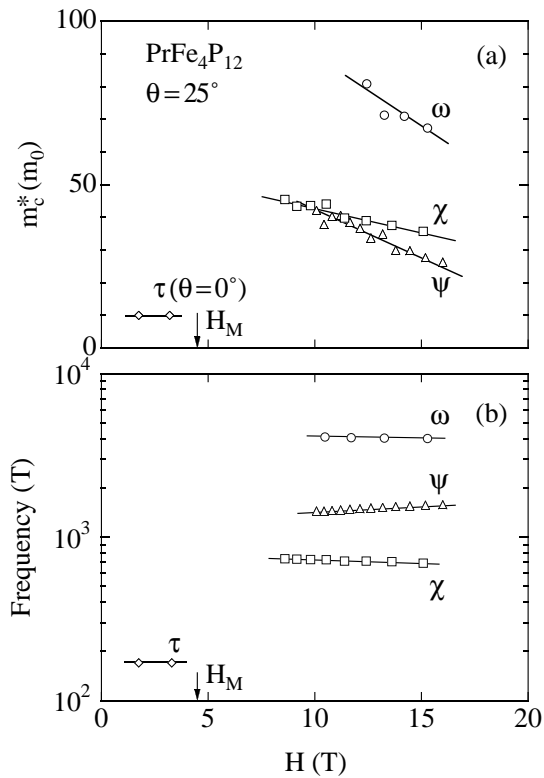}
\caption{(a) Field dependence of the $m^{\rm \ast}_{\rm c}$ and (b) dHvA frequencies for $\theta$=25$^\circ$ in PrFe$_4$P$_{12}$.}
\label{FieldDepMass&Freq}
\end{center}
\end{figure}

The dHvA experiments have directly proved the extraordinary enhancement of effective mass $m^{\rm \ast}_{\rm c}=81~{\it m}_{\rm 0}$ [see Fig.~\ref{FieldDepMass&Freq}~(a)] which gives the first unquestionable evidence for the heavy fermion state in Pr-based compounds; the $m^{\rm \ast}_{\rm c}$ is almost comparable with the reported heaviest masses of 120~$m_{\rm 0}$ for CeRu$_2$Si$_2$ and 105~$m_{\rm 0}$ for UPt$_3$.~\cite{Aoki,Kimura} That is consistent with the large $\gamma$-value and the Kondo-like features in the transport properties.
A noticeable feature in the $m^{\rm \ast}_{\rm c}$ is the large anisotropy as shown in Fig.~\ref{AngDepMass}; it gradually increases with increasing $\theta$ except for $\tau$-branch. Such a behavior is expected for a cylindrical FS along [001].
However, this is not the case, since the $m^{\rm \ast}_{\rm c}$ increases while the cross-sectional area decreases with increasing $\theta$ at least for $\psi$-branch as shown in Fig.~\ref{PrFe4P12dHvA_ang}. Such an anisotropic $m^{\rm \ast}_{\rm c}$, especially for the cubic system, suggests unusual anisotropic mass renormalization.

The most interesting subject is the origin of the mass enhancement, since the CEF ground state is believed to be a non-magnetic $\Gamma_{3}$. The mass enhancement in HFP and Kondo-like behaviors at high temperatures could be ascribed to the conventional Kondo effect associated with the magnetic CEF exited states; the proposed CEF level scheme is $\Gamma_{3}-\Gamma_{4}$,~\cite{Y_Aoki,Nakanishi} where the interaction between thermally induced magnetic moments and conduction electrons can bring about the Kondo effect due to the small excited energy ($\sim$10~K). 
The rearrangement of CEF levels in magnetic field stabilizing a magnetic ground state also could not be ruled out for the conventional Kondo effect in HFP.
In fact, as shown in  Fig.~\ref{FieldDepMass&Freq}~(a), the $m^{\rm \ast}_{\rm c}$ of the $\omega$, $\psi$, and $\chi$ branches are significantly suppressed by the magnetic field as was reported in typical Ce-based heavy fermion systems such as CeRu$_2$Si$_2$,~\cite{Aoki} CeB$_6$,~\cite{Joss} and CeCu$_6$.~\cite{Chapman}
The suppression of $m^{\rm \ast}_{\rm c}$ is consistent with the reduction of the $\gamma$-value by the magnetic field.~\cite{Y_Aoki}
However, the possibility of quadrupolar origin for the large mass in HFP still remains because the CEF excited $\Gamma_{4}$ state also has the quadrupole moments.
On the contrary, the effective mass of $\tau$-branch observed in LOP is insensitive against the magnetic field, which is also consistent with the almost constant $\gamma$-value in LOP.~\cite{Y_Aoki} These facts point to the non-magnetic origin of the large mass enhancement in this materials at least in LOP. The fluctuating quadrupolar moments which interact with conduction electrons could be responsible for the HF properties.

{\it 2)~Different FS topology with} LaFe$_4$P$_{12}$

Another unexpected feature in Fig.~\ref{PrFe4P12dHvA_ang} is a difference in the FS topology between PrFe$_4$P$_{12}$ and LaFe$_4$P$_{12}$. If the 4$f$-electrons are well localized in PrFe$_4$P$_{12}$, the FS should be close to that of LaFe$_4$P$_{12}$ at least in the HFP (or field induced ferromagnetic state) above $H_{\rm M}$. In fact, the FS of ferromagnetic NdFe$_4$P$_{12}$ was confirmed to be very close to that of LaFe$_4$P$_{12}$ except a small spin-splitting in the dHvA frequencies.~\cite{Sugawara_JPSJ} 
As shown in Fig.~\ref{FieldDepMass&Freq}~(b), apparent field dependence of the dHvA frequencies ($2\sim8\%$) is observed, at least for $\theta=0\sim25^\circ$. The frequencies for $\omega$- and $\chi$-branches decrease with increasing field, whereas the frequency for $\psi$-branch increases with increasing field. Such a field dependence of dHvA frequencies could be explained as due to the spin-splitting effect combined with the non-linear magnetization prosses,~\cite{Onuki,Nimori} if we assume that $\omega$- and $\chi$-branch originate from an up-spin state and $\psi$-branch originates from a down-spin state or vice versa. In the magnetic field, the spin degeneracy of conduction electrons is lifted and the each Landau energy level is split into the up- and down-spin states by the Zeeman effect. When the spin-splitting of the energy levels is proportional to the magnetic field and thereby the extremal cross-sectional area $S_{\rm F}$ of each spin state FS changes linearly with increasing field, the dHvA frequencies of each spin state coincide, giving the same $S_{\rm F}$ for zero field. On the other hand, in the magnetic materials, the conduction electrons have different Zeeman and exchange energies depending on the spin directions. In PrFe$_4$P$_{12}$, the 4$f$-moment is induced significantly by the magnetic field and the non-linear magnetization appears in HFP.~\cite{Y_Aoki} The exchange interaction between the 4$f$-electrons and conduction electrons brings about the non-linear field dependence of $S_{\rm F}$ which is a origin of the large field dependence of dHvA frequencies. Thus, a large spin-splitting effect could be an origin of the large difference in the dHvA branches between PrFe$_4$P$_{12}$ and LaFe$_4$P$_{12}$. However, we could obtain no conclusive evidence, since possible branches paired with the observed ones with different spin directions may not have been detected due to their heavy masses exceeding the present experimental sensitivity.~\cite{CompMass}

The question is why the Kondo effect (either magnetic or non-magnetic) which is quite unusual in Pr-compounds is realized in PrFe$_4$P$_{12}$. Crystallographically,
Pr-ions located within a large cage made of 12-P and 4-Fe ions in the filled skutterudite structure are under following special condition. Relatively large distance between Pr- and P-ions leads to the deeper 4$f$-level compared to the ordinary rare earth compounds,~\cite{Ishii} however, the large coordination number of 12 might fully compensate it and leads to even larger $c$-$f$ hybridization. {\it "The unique crystallographic structure could be an origin of such unusual feature in this compound as a Pr-compound through an enhanced c-f hybridization effect."}

In summary, we confirmed an enormously enhanced effective mass (the detected maximum value is $m^{\rm \ast}_{\rm c}=81~{\it m}_{\rm 0}$) in PrFe$_4$P$_{12}$. The insensitivity of mass against the magnetic field suggests non-magnetic origin of the mass enhancement at least in LOP whose order parameter to be also non-magnetic origin. PrFe$_4$P$_{12}$ may be a first Pr-based compound in which the quadrupolar Kondo effect coexists (or competes) with quadrupolar ordering. 

\begin{acknowledgments}
The authors are grateful to Prof.~H.~Harima, Dr.~D.~Aoki, Prof.~O.~Sakai and Prof.~M.~Kohgi for the helpful discussion. This work was supported by the Grant-in-Aid for Scientific Research from the Ministry of Education, Science, Sports and Culture of Japan.
\end{acknowledgments}


\begin{thebibliography}{99}
\bibitem{Table1}The dHvA frequencies and $m^{\rm \ast}_{\rm c}$ plotted in Fig.~\ref{PrFe4P12dHvA_ang} and listed in Table~\ref{table1} are average values over the field ranges $8-16$~T for  $\omega$-, $\psi$-, and $\chi$-branches, and $1.5-2.1$~T for $\tau$-branch.
\end{thebibliography}
\end{document}